\documentclass[aip,amsmath,apl,amssymb,reprint]{revtex4-2}
\usepackage{amsmath,amsfonts}
\usepackage{graphicx}
\usepackage[utf8]{inputenx}
\usepackage{etoolbox}
\usepackage[colorlinks=true,linkcolor=cyan,urlcolor=cyan,citecolor=cyan]{hyperref}
\usepackage[english]{babel}
\newcommand{\cm}{{cm$^{-1}$}}

\makeatletter
\def\@email#1#2{%
 \endgroup
 \patchcmd{\titleblock@produce}
  {\frontmatter@RRAPformat}
  {\frontmatter@RRAPformat{\produce@RRAP{*#1\href{mailto:#2}{#2}}}\frontmatter@RRAPformat}
  {}{}
}%
\makeatother
\begin{document}
\title{Contrasting properties of free carriers in $n$- and $p$-type Sb$_2$Se$_3$}

\author{F. Herklotz}
\email{frank.herklotz1@tu-dresden.de}
\affiliation{TU Dresden University of Technology, Institute of Applied Physics, 01062 Dresden, Germany}
\author{E. V. Lavrov}
\affiliation{TU Dresden University of Technology, Institute of Applied Physics, 01062 Dresden, Germany}
\author{T. D. C. Hobson}
\affiliation{Stephenson Institute for Renewable Energy, University of Liverpool, Liverpool L69 7ZF, UK}
\author{T. P. Shalvey}
\affiliation{Stephenson Institute for Renewable Energy, University of Liverpool, Liverpool L69 7ZF, UK}
\author{J. D. Major}
\affiliation{Stephenson Institute for Renewable Energy, University of Liverpool, Liverpool L69 7ZF, UK}
\author{K. Durose}
\affiliation{Stephenson Institute for Renewable Energy, University of Liverpool, Liverpool L69 7ZF, UK}

\date{\today}

\begin{abstract}
We report persistent photoconductivity in $p$-type Sb$_2$Se$_3$ single crystals doped with Cd or Zn, where enhanced conductivity remains for hours after illumination ceases at temperatures below $\sim$25~K. Comparative transport and infrared absorption measurements, including on $n$-type Cl-doped counterparts, reveal strong indications that hole transport in Sb$_2$Se$_3$ is more strongly affected by intrinsic carrier scattering than electron transport. These results point to a fundamental asymmetry in charge carrier dynamics and highlight the potential role of polaronic effects in limiting hole mobility in this quasi-one-dimensional semiconductor.
\end{abstract}

\maketitle

In recent years, antimony triselenide (Sb$_2$Se$_3$) has attracted considerable attention as a promising chalcogenide semiconductor for applications spanning photovoltaics,\cite{Chen_22_1} photoelectrochemical devices and photocatalysis,\cite{Yang_20_1} photodetectors,\cite{Zhai_10_1} thermoelectrics,\cite{Ko_16_1} batteries\cite{Ou_17} and phase-change materials\cite{Delaney_20_1}. A defining characteristic of Sb$_2$Se$_3$ is its quasi-one-dimensional crystal structure,\cite{Tideswell_57} comprising covalently bonded [Sb$_4$Se$_6$]$_n$ "nano-ribbons" that are held together by weak van der Waals interactions in the orthogonal directions. This structural anisotropy leads to strongly directional properties in carrier transport, optical absorption, and dielectric response, presenting unique opportunities for device applications—provided the underlying physical mechanisms are well understood.

Among its various applications, Sb$_2$Se$_3$ has shown particular promise as a photovoltaic absorber material, with reported power conversion efficiencies exceeding 10\%, despite the relatively limited research effort compared to mainstream thin-film materials such as CdTe or CIGS.\cite{Dale_23} However, further improvement is currently constrained by a low open-circuit voltage, commonly attributed to factors such as low carrier concentration and recombination losses mediated by bulk or interface defects.

In polar semiconductors, strong electron–phonon interactions can lead to the formation of small and large polarons,\cite{Holstein_59_2,Emin_13} where charge localization distorts the lattice and hampers mobility. While such effects have been proposed in Sb$_2$Se$_3$ and its analog Sb$_2$S$_3$,\cite{Tao_22,Wang_19,Yang_19,Grad_21,Chong_14,Liu_23} their actual impact on carrier transport remains debated. Unlike defect-assisted recombination, which may be mitigated by improved synthesis or interface engineering, polaronic self-trapping is inherently difficult, if not impossible, to eliminate.

Recent advances in bulk crystal growth have enabled the synthesis of $n$-type Sb$_2$Se$_3$ single crystals via Cl-doping, achieving room-temperature carrier concentrations exceeding $10^{16}$~cm$^{-3}$ and conductivities on the order of 0.1~$\Omega^{-1}$cm$^{-1}$.\cite{Hobson_20_2} These samples have enabled detailed studies of charge transport, including investigations of free-carrier absorption\cite{Herklotz_24_1} and persistent photoconductivity (PPC),\cite{Herklotz_25_2} which have shown that electrons in Sb$_2$Se$_3$ generally exhibit band-like transport, characterized by relatively weak electron–phonon coupling. This suggests that small-polaron formation is not a limiting factor for electron mobility in this material.

In this Letter, we present a comparative study of photoconductivity in $p$-type and $n$-type Sb$_2$Se$_3$, with a particular emphasis on the role of carrier–scattering mechanisms. Using Sb$_2$Se$_3$ single crystals doped with Cd and Zn to achieve highly $p$-type behavior, we observe persistent photoconductivity at cryogenic temperatures, with lifetimes extending well beyond the duration of optical excitation. When compared to results from $n$-type Cl-doped crystals, our findings reveal a marked asymmetry in free carrier absorption between holes and electrons, with the former exhibiting stronger indications of polaronic behavior. These results suggest that intrinsic scattering mechanisms may play a greater role in limiting hole mobility, with implications for the design and optimization of Sb$_2$Se$_3$-based optoelectronic devices.

Single crystals were grown by vertical Bridgman melt-growth, either undoped or doped in-growth with Cl, Cd, Zn, Sn, or O.\cite{Hobson_20_3,Hobson_20_2,Shalvey_25} The resulting ingots (typically up to 4~mm in diameter, 1–3~cm in length) were cleaved, cut, and mechanically polished to produce $b$-oriented slices (in the \textit{Pbnm} space group notation), with front-face dimensions of $\sim$3~$\times$~3~mm$^2$.

Temperature-dependent current–voltage ($I$–$V$) measurements were performed using a Keithley 2601A source-measure unit in a two-point probe configuration. Electrical contacts were aligned along the crystallographic $c$-axis within the $b$-face of each sample. Ohmic contacts for the $n$-type Cl-doped crystals were fabricated by thermal evaporation of In, whereas for the nominally undoped and Cd-, Zn-, O-, and Sn-doped samples, Au contacts were evaporated, with typical contact areas of $\sim$0.5~$\times$~0.5~mm$^2$. Gold wires were used to establish electrical connections.

Infrared (IR) absorption measurements were conducted using a Bomem DA3.01 Fourier-transform infrared spectrometer, equipped with a globar source and interchangeable beam splitter–detector combinations: 3~$\mu$m Mylar or KBr beam splitters, and either a silicon bolometer or a mercury cadmium telluride (MCT) detector. The IR beam direction $k$ was oriented along the sample’s $b$-axis. Polarization-resolved measurements were enabled by a wire-grid polarizer mounted on a KRS-5 substrate. The spectral resolution was set to 1--2~cm$^{-1}$.

All temperature-dependent $I$–$V$ and IR absorption measurements were conducted with the samples mounted inside a continuous-flow helium cryostat equipped with polypropylene outer and polyethylene inner windows. For photoexcitation, samples were illuminated using a 150~W xenon short-arc (XBO) lamp. The illumination path included KBr outer windows (bandgap: $\sim 4.3$ eV) and ZnSe inner side windows (bandgap: $\sim 2.7$ eV) of the cryostat to ensure broadband transmission while minimizing unwanted absorption effects.

\begin{figure}[t]
    \includegraphics[width=0.48\textwidth]{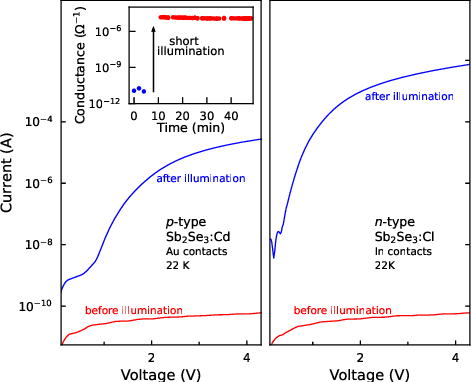}
    \caption{Semi-logarithmic $I$--$V$ characteristics of $p$-type Cd-doped (left panel) and $n$-type Cl-doped (right panel) Sb$_2$Se$_3$ single crystals measured at low temperature ($\sim$22~K), before and after 1-minute illumination with an XBO lamp. The inset shows the temporal evolution of the sample conductance (measured at 4~V) for the Cd-doped crystal, highlighting the persistent nature of the photoconductivity.}
    \label{IVcurves}
\end{figure}

Figure~\ref{IVcurves} compares the semi-logarithmic $I$--$V$ characteristics of Ohmic contacts on Cd-doped (left) and Cl-doped (right) Sb$_2$Se$_3$ single crystals, measured at approximately 22~K following cooldown in darkness. The majority carrier type in each sample was independently confirmed using (i) the hot-probe technique~\cite{Golan_06_1} and (ii) Hall effect measurements. These analyses established that the Cd-doped crystals are $p$-type, with a room-temperature hole concentration of approximately $4 \times 10^{15}$~cm$^{-3}$ and an isotropically averaged hole mobility of 1--2~cm$^2$/Vs. In contrast, the Cl-doped crystals are $n$-type, with both electron concentration and mobility roughly an order of magnitude higher.\cite{Hobson_20_2} A detailed discussion of Hall measurements on Cd-doped samples will be provided elsewhere.\cite{Shalvey_25} It is also important to note that, due to the differing Fermi level positions in $p$- and $n$-type Sb$_2$Se$_3$, distinct metallization schemes were employed to ensure Ohmic behavior: Au was thermally evaporated onto the $p$-type (Cd-doped) samples, whereas In contacts were used for the $n$-type (Cl-doped) crystals.

The red curves in Fig.~\ref{IVcurves} represent $I$--$V$ measurements recorded immediately after cooling, revealing currents close to the instrument's noise floor in both samples. Following 1 minute of illumination with a broadband XBO lamp (power density $\sim$10~mW/cm$^2$) at 22~K, both samples exhibited a significant increase in conductivity, as indicated by the blue curves. The enhancement typically initiates within seconds and approaches a quasi-steady-state under continued illumination over several minutes. 

The inset in the left panel of Fig.~\ref{IVcurves} shows the time-dependent conductance of the Cd-doped crystal, measured at 4~V, before and after illumination. The data confirm that the photo-induced increase in current endures after the excitation light is turned off, indicating the presence of persistent photoconductivity in these $p$-type samples. A comparable PPC response was observed in the $n$-type Cl-doped crystals, consistent with previous reports.\cite{Herklotz_25_2}

The observation of persistent photoconductivity at low temperature was consistently found across all Cd- and Cl-doped Sb$_2$Se$_3$ crystals studied. In addition, PPC was also observed in Zn-doped samples, which showed very similar electrical characteristics to the Cd-doped ones. In contrast, PPC was absent in all nominally undoped or oxygen- and tin-doped crystals studied in this work. These samples exhibited room temperature conductivities at least three orders of magnitude lower than those of the Cl-, Cd-, and Zn-doped samples. These findings strongly suggest that the observed persistent photoconductivity originates from carriers introduced via intentional Cl, Cd, or Zn doping.
 
\begin{figure}[t]
    \includegraphics[width=0.48\textwidth]{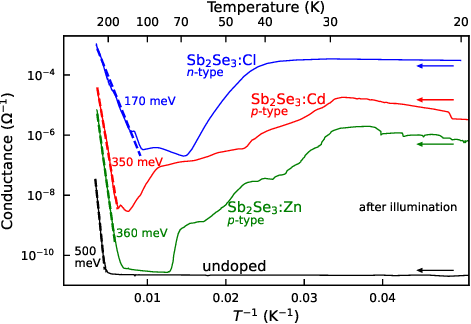}
\caption{Arrhenius plots of sample conductance during heat-up for Cl-, Cd-, Zn-, and nominally undoped Sb$_2$Se$_3$ crystals after a 1-minute illumination with a XBO lamp at 18~K. Measurements were performed in the dark under a 5~V bias with a heating rate of approximately 2~K/min.}
\label{heatup}
\end{figure}

Figure~\ref{heatup} presents Arrhenius plots of the conductance during heat-up (measured at 5~V bias) for Cl-, Cd-, Zn-, and nominally undoped Sb$_2$Se$_3$ samples following 1-minute illumination with an XBO lamp at 18~K. All measurements were performed in the dark while heating the samples at a rate of approximately 2~K/min.

As shown in the figure, the undoped sample exhibits no measurable photoconductivity at low temperatures, with conductance remaining at the noise level. Upon heating above $\sim$200~K, the conductance increases exponentially with an activation energy of approximately 500~meV. This behavior closely matches that of the dark conductivity (not shown), confirming a purely thermally activated conduction mechanism in the absence of significant photocarrier contribution.

In the $n$-type Cl-doped sample, the PPC remains nearly constant in the range between 18 and 40~K. Above 40~K, the photoconductivity declines rapidly, demonstrating thermal instability of the photoexcited carriers at this temperatures. A weak, broad conductivity peak observed between 70 and 110~K is tentatively attributed to thermally released carriers from deep traps populated during prior illumination. Above 110~K, the conductance aligns with the dark conductivity curve (not shown), indicating a transition to thermally activated conduction with an activation energy of $\sim$170~meV. This value is consistent with prior studies\cite{Herklotz_24_1,Herklotz_25_2} and is attributed to the thermal ionization energy of substitutional Cl$_\mathrm{Se}$ donors.\cite{Hobson_20_2,Stoliaroff_20}

By contrast, the heat-up behavior of the Cd- and Zn-doped $p$-type samples differs markedly from the Cl-doped case but shows strong similarity between each other. In both cases, the PPC increases between 18 and 28~K, likely due to enhanced carrier mobility at elevated temperatures. Beyond this point, the photoconductivity decreases gradually, exhibiting a series of weak, broad features that are attributed to thermally stimulated currents originating from deep trap states. At higher temperatures, the conductance follows Arrhenius-like behavior with activation energies of approximately 350~meV (Cd) and 360~meV (Zn), potentially corresponding to the thermal ionization of the respective acceptor levels. The qualitative similarity in PPC behavior between the Cd- and Zn-doped $p$-type crystals, and its distinct difference from the $n$-type Cl-doped case, suggests a fundamental asymmetry in the photoconductive behavior of holes and electrons in Sb$_2$Se$_3$. 

\begin{figure}[t]
    \centering
    \includegraphics[width=0.48\textwidth]{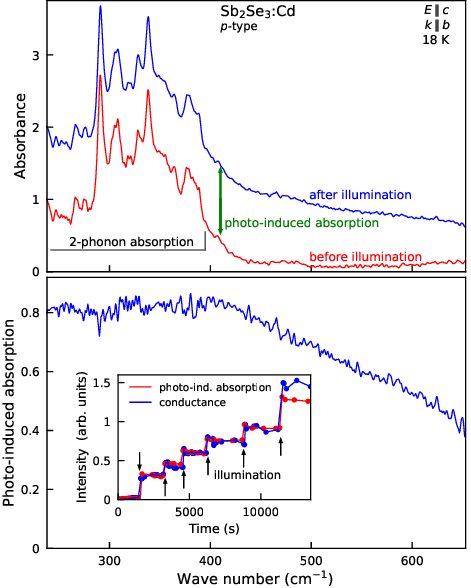}
    \caption{Top panel: IR absorbance spectra obtained on a Cd-doped Sb$_2$Se$_3$ single crystal after cooling down to 18 K in dark, measured before (red) and after (blue) 10 s of illumination using an XBO light source. Bottom panel: "differential" absorbance spectra with the spectra measured before and after illumination cross-subtracted. Inset: Temporal evolution of conductance (blue) and photo-induced absorption (red) of the Cd-doped Sb$_2$Se$_3$ sample, measured during a series of short illumination steps at 18 K after cooling down in dark.}
    \label{PIA}
\end{figure}

To further investigate the origin of the persistent photoconductivity observed in Cd-doped Sb$_2$Se$_3$, we performed IR absorption measurements. Figure~\ref{PIA} displays absorption spectra recorded for such a sample, with the probe beam propagating along the $b$ axis and the electric field vector of the light polarized parallel to the $c$ axis—aligned with the material's quasi-one-dimensional [Sb$_4$Se$_6$]$_n$ nano-ribbons.

The upper panel compares spectra measured at $\sim$18~K in the dark (red) and after 10~s of illumination using the XBO light source (blue). The lower panel shows the differential spectrum obtained by subtracting the dark spectrum from the illuminated one. The emergence of photo-induced absorption (PIA) is clearly visible following illumination. Strong absorption bands between 240 and 400~{\cm}, attributable to two-phonon transitions, are also observed, though their detailed analysis lies beyond the scope of this study.

The inset in the lower panel plots the PIA intensity (integrated over 240--650~{\cm}) alongside the sample's electrical conductance (@ 5 V), measured during a series of short illumination steps. Both quantities are normalized for comparison. The stepwise increase in both PIA and photoconductance confirms their strong correlation and indicates a common physical origin, most likely related to the generation of free charge carriers.

\begin{figure}[t]
    \centering
    \includegraphics[width=0.48\textwidth]{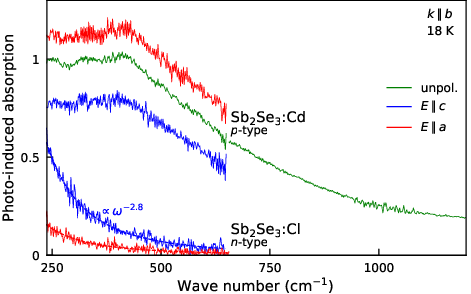}
    \caption{Polarization-resolved differential PIA spectra of Cd- and Cl-doped Sb$_2$Se$_3$ single crystals, obtained by subtracting the dark spectra from those measured after illumination. Blue and red curves correspond to light polarized along the $c$ and $a$ axes, respectively, while green curves represent measurements with unpolarized light. Data below 630~{\cm} were recorded using a Mylar beam splitter and bolometer detector combination; for frequencies above 630~{\cm}, a KBr beam splitter and MCT detector setup was used.}
    \label{polarized}
\end{figure}

Figure~\ref{polarized} presents polarization-resolved differential PIA spectra of Cd-doped ($p$-type) and Cl-doped ($n$-type) Sb$_2$Se$_3$ crystals obtained after illumination. The blue and red curves correspond to light polarized along the crystallographic $c$ and $a$ axes, respectively. The green curves were recorded with unpolarized light.

A key observation for the $n$-type Cl-doped sample is its spectral behavior: the PIA exhibits a power-law dependence $\alpha \propto \omega^{-n}$ with $n \approx 2.8$. This is characteristic of free-carrier absorption in semiconductors, where the exponent $n$ typically ranges from 1.5 to 3.5 depending on the dominant carrier scattering mechanism.\cite{Fan_56,Meyer_58,Visvanathan_60,Gurevich_62} Additional support for this interpretation comes from the polarization dependence: absorption is about four times stronger for light polarized along the $c$ axis than the $a$ axis. This anisotropy reflects the quasi-one-dimensional crystal structure of Sb$_2$Se$_3$, which dictates the anisotropy of the Fermi surface and conduction band effective masses,\cite{Wang_22_1,Wang_22_2} and aligns with earlier reports of anisotropic free-carrier absorption in this material.\cite{Herklotz_24_1,Herklotz_25_2}

In contrast, the $p$-type Cd-doped samples exhibit qualitatively different spectral features. For both polarization directions, the PIA signal shows a broad plateau extending up to approximately 420~{\cm}, close to the upper boundary of the two-phonon spectral region (cf.~Fig.~\ref{PIA}). At higher wavenumbers, the absorption gradually decreases without exhibiting a clear $\omega^{-n}$ dependence. Additionally, the polarization anisotropy is reversed compared to the $n$-type samples, with the PIA slightly stronger for light polarized along the $a$ axis than along the $c$ axis. Notably, the overall intensity of the photo-induced absorption in the $p$-type sample is substantially higher than in the $n$-type, despite the photoconductivity being roughly two orders of magnitude lower (see Figs.~\ref{IVcurves} and~\ref{heatup}).

These contrasting behaviors point to a fundamental difference in charge carrier scattering mechanisms between $n$- and $p$-type Sb$_2$Se$_3$, possibly linked to differences in electron- and hole-phonon coupling strengths and polaronic effects. The significantly lower carrier mobility observed in our $p$-type crystals compared to their $n$-type counterparts supports this hypothesis. Further analysis and complementary experimental data are needed to clarify the underlying mechanisms.

In summary, we report persistent photoconductivity in $p$-type Sb$_2$Se$_3$ single crystals doped with Cd or Zn at temperatures below 25~K. Comparative electrical transport and infrared absorption studies of $p$-type and $n$-type (Cl-doped) samples reveal a pronounced asymmetry in charge carrier behavior. Hole transport shows stronger signatures of carrier scattering than electron transport. This asymmetry, supported by spectral and polarization-resolved infrared data, highlights distinct mechanisms influencing electron and hole mobility in this quasi-one-dimensional semiconductor.

\begin{acknowledgments}
This work was funded by the Deutsche Forschungsgemeinschaft (Grant No. LA 1397/21) as well as the Engineering and Physical Sciences Research Council grants EP/W03445X/1 and EP/M024768/1.
\end{acknowledgments}

\section*{Data Availability Statement}
The data that support the findings of this study are available from the corresponding author upon reasonable request.

%

\end{document}